\documentclass[twocolumn,floats,floatfix,showpacs,amssymb,prd,superscriptaddress,nofootinbib]{revtex4}

\usepackage{graphicx,epsf, epsfig, amssymb}
\usepackage{bm}
\usepackage{longtable}
\usepackage{color}
\usepackage{amsfonts,amsmath,amssymb,mathrsfs,times}

\def\be{\begin{equation}}
\def\ee{\end{equation}}
\def\beq{\begin{eqnarray}}
\def\eeq{\end{eqnarray}}

\begin{document}

\title{Testing  the Cosmic Censorship Conjecture with point particles: the effect of radiation reaction
and the self-force}

\pacs{04.70.Bw, 04.20.Dw}

\author{Enrico Barausse} \email{barausse@umd.edu}
\affiliation{Department of Physics, University of Maryland, College
Park, MD 20742, USA}

\author{Vitor Cardoso} \email{vitor.cardoso@ist.utl.pt}
\affiliation{CENTRA, Departamento de F\'{\i}sica, Instituto Superior T\'ecnico, Universidade T\'ecnica de Lisboa - UTL,
Av.~Rovisco Pais 1, 1049 Lisboa, Portugal}
\affiliation{Department of Physics and Astronomy, The University of Mississippi, University, MS 38677, USA}

\author{Gaurav Khanna} \email{gkhanna@umassd.edu}
\affiliation{Physics Department, University of Massachusetts
  Dartmouth, North Dartmouth, MA 02747, USA}

\begin{abstract}
A classical thought-experiment to destroy black holes was envisaged by Wald in 1974: it consists of 
throwing particles with large angular momentum into an extremal black hole, checking whether their capture
can over-spin the black hole past the extremal limit and create a naked singularity.
Wald showed that in the test-particle limit, particles that would be otherwise capable of producing naked singularities are simply scattered.
Recently Jacobson and Sotiriou showed that if one considers instead a black hole that is almost, but not exactly extremal,
then in the absence of backreaction effects particle capture could indeed over-spin the spacetime above the Kerr limit.
Here we analyze back-reaction effects and show that for some of the trajectories giving rise to naked singularities, 
radiative effects can be neglected. However, for these orbits the conservative self-force is important, 
and seems to have the right sign to prevent the formation of naked singularities.
\end{abstract} 

\maketitle

\section{Introduction}

The appearance of singularities as solutions of a theory's field equations often signals the breakdown of that theory
in some regime. In classical electrodynamics, for instance, the field of a point-like charge (the basic object of the theory) diverges at the particle's location. This singularity can be resolved only within quantum electrodynamics. Newton's theory also exhibits singularities at the location of point masses, and these singularities remain in General Relativity. In 1939, Oppenheimer and Snyder~\cite{Oppenheimer:1939ue} showed 
that under the assumption of perfect spherical symmetry, a sufficiently massive star that runs out of thermonuclear fuel
will undergo indefinite collapse to a point, where the curvature invariants diverge.
In fact, divergent curvature invariants as the outcome of collapse are a generic feature of General Relativity, even in the 
absence of spherical symmetry, as shown by Penrose in 1965~\cite{sing_theorem}. It is expected that a quantum theory of gravity will 
be able to smoothen and resolve this singular behavior.

Fortunately, the equations of General Relativity seem to contain a self-protection mechanism (from the unpredictability of naked singularities), 
hiding curvature singularities within event horizons and  making them invisible to outside observers and therefore to experiments.
This is known as the ``Cosmic Censorship Conjecture'', and was proposed by Penrose in 1969~\cite{CCC}. While several counterexamples
to this conjecture are known, they either rely on using certain equations of state beyond their range of validity 
(such as the  pressureless matter equation of state used by Ref.~\cite{dust}\footnote{Note that pressureless matter 
gives rise to shell-crossing singularities even in flat-spacetime evolutions, but these singularities are unphysical because
the pressureless equation of state does not hold at infinite densities.}); or they are non-generic in nature 
(\textit{i.e.,} they only happen for very specific sets of initial data: see Ref.~\cite{klein_gordon}); or they are staged in higher-dimensional spacetimes~\cite{Lehner:2010pn}. 
Therefore, while a general proof of the Cosmic Censorship Conjecture remains elusive, all existing evidence points at some version of it being true
in generic 4-dimensional, asymptotically flat spacetimes~\cite{Wald:1997wa}. This is reassuring because it means that a
quantum theory of gravity is not needed to understand the astrophysics of gravitationally-collapsed objects\footnote{The only exception is given
by Hawking radiation~\cite{hawking}. However, that is a semiclassical effect, and is too weak to be relevant in most astrophysical situations.}, since all curvature singularities
would be cloaked behind an event horizon and would therefore be inaccessible to external observers.

Because, according to the Cosmic Censorship Conjecture, curvature singularities are generically 
hidden behind black-hole event horizons, the most natural way
to search for naked singularities is to look for instabilities of the Kerr geometry, which is the most general stationary vacuum black-hole solution of Einstein's equations in a four-dimensional,
asymptotically flat spacetime~\cite{Kerr:1963ud}. 
At the linearized level, all existing evidence points to the stability of the Kerr event horizon\footnote{The Cauchy horizon of the Kerr metric, which lies
inside the event horizon, is instead known to be unstable both at the linear and nonlinear level~\cite{cauchy}.}:

\noindent (i) the Kerr metric outside the event horizon is perturbatively 
stable against exponentially growing modes of massless fields~\cite{Whiting:1988vc}. 
The Kerr exterior geometry is unstable against massive scalar 
fields~\cite{Damour:1976kh,Zouros:1979iw,Detweiler:1980uk,Cardoso:2005vk,Strafuss:2004qc,Dolan:2007mj}, but the 
instability is thought to extract rotational energy from the geometry, and does not in principle destroy the horizon. 

\noindent (ii) Because naked singularities appear in the Kerr geometry when the angular momentum of the spacetime is above the Kerr bound, 
\textit{i.e}. for $cJ/GM^2 > 1$, a simple gedanken experiment was proposed by Wald~\cite{wald} to try to create naked singularities. 
This consisted of throwing point particles having large angular momentum into an extremal Kerr black hole. 
In the test-particle approximation (\textit{i.e.} at the lowest order in the mass of the particle), Wald finds 
that particles that would over-spin the geometry past the Kerr bound are not captured, but are scattered. 
This conclusion generalizes to a wide variety of backgrounds~\cite{BouhmadiLopez:2010vc}.

On the other hand, Kerr spacetimes that present $cJ/GM^2 > 1$, and which therefore contain naked singularities, are linearly 
unstable~\cite{unstable_naked}. 
Thus, a consistent picture emerges,  suggesting that black holes are stable and do not give rise to naked singularities. 

At the full nonlinear level, all numerical-relativity evolutions of Einstein equations that asymptote to the 
Kerr geometry at late times  are consistent with this scenario, 
suggesting stability of the Kerr spacetime~\cite{Sperhake:2009jz,shibata,Lovelace:2010ne,kesden1,finalspin,bkl,fau,rit}. 
In particular, a possible generalization of Wald's process to comparable mass ratios has been studied in Refs.~\cite{Sperhake:2009jz,shibata},
which considered high-energy collisions between two comparable-mass non-spinning black holes.
Refs.~\cite{Sperhake:2009jz,shibata} found no evidence of formation of naked singularities, 
because either the black holes are simply scattered to infinity, 
or the full nonlinear equations make the system radiate enough angular momentum to form a single black hole. 
(More specifically, for finely tuned impact parameters, the black holes spend a finite 
amount of time in zoom-whirl orbits and radiate the excess angular momentum, eventually being ``allowed'' to merge.)

Similarly, Abrahams~\textit{et al.}~\cite{abrahams_etal} studied the collapse
of axisymmetric tori made of collisionless matter, and found that 
tori with $cJ/GM^2<1$ (``sub-Kerr'' configurations) collapse to Kerr black holes,
while ``supra-Kerr'' tori with $cJ/GM^2>1$ collapse to new equilibrium configurations.
The collapse of rotating stars has also been investigated by a number of authors~\cite{nakamura,sato_nakamura,stark_piran,shibatabis,duez,giacomazzo}. 
These studies show that the collapse of sub-Kerr stellar models produces Kerr black holes, while the
collapse of supra-Kerr models does not give rise to naked singularities. For instance, Ref.~\cite{giacomazzo}
shows that the collapse of a supra-Kerr differentially rotating polytropic star produces a torus, which then fragments in
clumps that merge again, forming a bar and eventually a stable axisymmetric configuration.

Recently however, Jacobson and Sotiriou (JS)~\cite{Jacobson:2009kt} (building on Refs.~\cite{previous}) have 
re-considered Wald's process, but using almost extremal black holes, rather than exactly extremal ones.
Surprisingly, JS showed that this change is enough to allow test-particles 
with dangerously large angular momentum to be captured, over-spinning the black hole and creating naked singularities.
However, the particles considered by JS need to have  a finite mass-energy to overspin the black hole, 
and in this sense the situation that they consider is intermediate between Wald's original construction for test particles,  
and the full nonlinear analysis of Refs.~\cite{Sperhake:2009jz,shibata} for comparable masses. Therefore,
as acknowledged by JS themselves, a test-particle analysis such as that of Wald's gedanken experiment is likely not
adequate for this scenario, because it neglects the  conservative and dissipative self-force, which may be  important~\cite{hod}.

Here we expand on our previous {\it Letter}~\cite{letter}, and show that the dissipative self-force (equivalent to radiation reaction, 
\textit{i.e.}~the energy and angular momentum losses through gravitational waves) can prevent the formation of naked 
singularities only for some of JS's orbits. However, we will show that for \textit{all} these orbits 
the conservative  self-force is comparable to the terms giving rise to naked singularities, 
and should therefore be taken into account. 

In particular, in Sec.~\ref{orbits} we review the orbits that JS identified as giving rise to naked singularities. In Sec.~\ref{fluxes_sec}
we show how to calculate the gravitational-wave fluxes produced by these orbits, both analytically (Sec.~\ref{analytics}) and numerically
(Secs.~\ref{numerics} and \ref{extrapolation}). A comparison between the analytical and numerical fluxes is presented in Sec.~\ref{comparison},
while in Sec.~\ref{conservative_sf_section} we analyze the effect of the conservative self-force on the JS process. In Sec.~\ref{concl} we present
our conclusions. 
Hereafter we set $G=c=M=1$.

\section{The overspinning orbits of Jacobson \& Sotiriou}
\label{orbits}
Let us consider a Kerr black hole with spin parameter $a\equiv J/M^2=1-2 \epsilon^2$, with $\epsilon \ll 1$, 
and a non-spinning test-particle with energy $E$, angular momentum $L$ and mass $\mu$.
By adopting suitable length units, we can set the mass $M$ of the black hole  to 1 without loss of generality,
and it therefore follows that in order for the test-particle approximation to be valid it must be $\mu\ll 1$
as well as $E\ll 1$ and $L\ll1$.

Neglecting the loss of energy and angular momentum through gravitational waves 
(which is equivalent to the so-called dissipative self-force) and the conservative self-force (\textit{i.e.,} the modifications of the effective potential for the particle's motion due to the small but finite mass ratio of the system), the particle moves on a geodesic of the background
Kerr spacetime. JS then impose \textit{(i)} that this geodesic orbit fall into the black hole,  and \textit{(ii)} 
that the Kerr black hole be spun up past the
extremal limit when the particle is captured, \textit{i.e.} that the final spin $a^{\rm JS}_{\rm f}=(a+L)/(1+E)^2$ be larger than 1.
Condition  \textit{(i)} implies an upper limit $L<L_{\max}$ on the angular momentum, because particles with large angular momenta
are simply scattered, while condition \textit{(ii)} implies a lower limit  $L>L_{\min}$ on the angular momentum,
because the particle needs to transfer a sufficient amount of angular momentum to the black hole in order to spin it up past the extremal limit.
In particular, one finds 
\beq
L_{\min}=2 \epsilon^2 + 2 E + E^2<L< L_{\max}= (2 + 4 \epsilon) E\,.\label{limitsJ}
\eeq
In order for orbits satisfying both  \textit{(i)}  and  \textit{(ii)} to exist, one has then to impose  $L_{\max}>L_{\min}$, which
yields  
\beq
E_{\rm min}= (2 - \sqrt{2})\epsilon<E< E_{\max}= (2 + \sqrt{2})\epsilon\,.\label{limitsE}
\eeq
This interval is indeed finite when $\epsilon>0$, while it shrinks to zero when $\epsilon=0$ (\textit{i.e.,} when $a=1$). 
This confirms Wald's results that an extremal black hole cannot be spun up past the extremal limit by the capture of test particles.
Lastly, JS noticed that intervals \eqref{limitsJ} and \eqref{limitsE} contain both {\it bound} orbits (\textit{i.e.} 
orbits that start with zero radial velocity at finite radius) and {\it unbound} orbits (\textit{i.e.} orbits that start from infinity).

Parameterizing the above intervals as
\begin{align}
&E=E_{\min}+x (E_{\max}-E_{\min})= E_{\min}+ 2 x \sqrt{2} \epsilon\label{Ex}\\
&L=L_{\min}+y (L_{\max}-L_{\min})=L_{\min} + 8 y \epsilon^2 (1-x) x\label{Ly}
\end{align}
with $0<x<1,\,0<y<1$, the final spin is
\be
a^{JS}_f=\frac{a+L}{(1+E)^2}= 1+8 \epsilon^2 (1-x) x y+{\cal O}(\epsilon^3)>1\,,\label{af}
\ee
which is indeed larger than 1 by terms that are quadratic in $\epsilon$. It is clear, however, that the inclusion 
of the gravitational-wave losses of energy and angular momentum,  $E_{\rm rad}$ and $L_{\rm rad}$, can in principle
affect JS's analysis by changing the prediction \eqref{af}
for the final spin to
\be
a_f=
1+8 \epsilon^2 (1-x) x y+2E_{\rm rad}-L_{\rm rad}+{\cal O}(\epsilon^3)\,.
\label{radcorrections}
\ee
In order to analyze the magnitudes of  $E_{\rm rad}$ and $L_{\rm rad}$ relative to the
spin-up term in  \eqref{radcorrections}, we restrict our attention on  {\it unbound} orbits,\footnote{As we 
mentioned, JS also considered bound orbits, falling 
into the Kerr black hole from a Boyer-Lindquist radius $r=r_{\rm hor}+{\cal O}(\epsilon)$ 
($r_{\rm hor}$ being the horizon's radius). However, these orbits pose a problem,
as we will show later, 
because the distance to the horizon is comparable to the particle's
\textit{minimum} attainable size $\max(E,\mu)\gtrsim \epsilon$, so finite-size effects should be taken into account.}
and following JS we assume $E/\mu\gg1$ and $L/\mu\gg1$ (null orbits) to further simplify our analysis. These relativistic 
orbits are characterized by the impact parameter $b=L/E$ alone.
From Eqs.~\eqref{limitsJ} and~\eqref{limitsE}, it follows that 
JS's orbits have $L=bE$, with 
$b=2 + 4 \epsilon [1 - 2 x (x - 1) (y - 1)/(2 + \sqrt{2} (2 x - 1))]$.
Varying $x$ and $y$ between 0 and 1, one obtains 
$b=2+\delta\epsilon$ with $2 \sqrt{2}<\delta<4$. 
However, because $b_{\rm ph}=2+2 \sqrt{3}\epsilon+{\cal O} (\epsilon^2)$ is the impact parameter
of the circular photon orbit (or ``light-ring''),  only orbits with $2 \sqrt{2}<\delta< 2 \sqrt{3}$ are unbound.

When $\delta\approx2 \sqrt{3}$, the particle has an impact parameter very close to that 
of the light-ring, and therefore orbits around the light-ring many times before plunging into the black hole. 
The emission of gravitational waves should be important for these orbits (because the gravitational-wave fluxes
will be proportional to the number of cycles at the light-ring), and could prevent the formation of naked singularities 
or at least invalidate
JS's analysis. In fact, for $\delta$ {\it arbitrarily} close to $2 \sqrt{3}$, 
the particles would orbit around the light-ring an arbitrarily large number of times, 
and gravitational-wave emission \textit{must} be important~\cite{Berti:2009bk}. 
We will show, however, that this is \textit{not} true for all of JS's orbits but only for a subset of them. Nevertheless, we
will also show that for all of JS's orbits, the conservative self-force is always important and seems to have the right sign to
prevent the formation of naked singularities.

 To estimate the magnitudes of  $E_{\rm rad}$ and $L_{\rm rad}$, we first compute
the number of cycles $N_{\rm cycles}$ described by the relativistic particle at the light-ring using the geodesics equations. 
As we have mentioned,
$E_{\rm rad}$ and $L_{\rm rad}$ are expected to be proportional to $N_{\rm cycles}$, because the plunge from infinity to the light ring 
and that from the light-ring to the horizon are not expected to produce significant amounts of gravitational waves, since they happen on a
dynamical timescale.

 The geodesic equations for null equatorial orbits read
 \begin{align}
 &\left(dr/d\lambda\right)^2= V_r(r)\,,\quad d\phi/d\lambda=V_\phi(r)\,,\\
 &V_r(r)= r \left[r^3-4 a b-b^2 (r-2)+a^2 (r+2)\right],&
 \label{rdot}\\
 &V_\phi(r)=
   \frac{[2 a+b (r-2)] r}{a^2+(r-2) r},&
 \label{phidot}
 \end{align}
 where $\lambda$ is a non-affine parameter ($d\lambda=d\lambda_{\rm affine}E/r^2$).
 For $b=b_{\rm ph} (1-k)$, with $k\ll\epsilon \ll1$, the radial potential presents a minimum $V_r^{\min}$ at $r=r_{\min}$:
 \begin{align}
 &V_r^{\min}\approx 8 k\epsilon/\sqrt{3} +{\cal O}(k \epsilon^2)+{\cal O}(k^2)\,,\\
 &r_{\min}\approx r_{\rm ph} -4 k (1+\sqrt{3} \epsilon)/3+{\cal O}(k \epsilon^2)+{\cal O}(k^2)\,.
 \end{align}
 Near this minimum one has
 \begin{align}
 &\!\!\!\!V_r(r)\!=\!V_r^{\min}\!\!+\frac12 V_r''(r_{\min}) (r-r_{\min})^2\!\!+{\cal O} (r-r_{\min})^3\,\\
 &\!\!\!\!V_\phi(r)=\frac83+\frac{\sqrt{3}}{2 \epsilon}+{\cal O}\left( \epsilon\right)+{\cal O}\left(k\right)+{\cal O} (r-r_{\min})\,,
 \end{align}
 where $V_r''(r_{\min})=6+{\cal O}\left( \epsilon\right)+{\cal O}\left(k\right)$. 
 Therefore, we get
 \be
 \frac{d\phi}{dr}\approx \left(\frac83+\frac{\sqrt{3}}{2 \epsilon}\right)\left[\frac{8}{\sqrt{3}} k\epsilon+3 (r-r_{\min})^2\right]^{-1/2}\,.
 \ee
 Integrating from $r_{\min}-\Delta r_1$ to $r_{\min}+\Delta r_2$, with $\Delta r_{1,2}\gg k \epsilon$, the number of cycles near the minimum is
 \begin{equation}
 \label{cycles}
 N_{\rm cycles} \approx \int^{r_{\min}+\Delta r_2}_{r_{\min}-\Delta r_1} \frac{d\phi}{dr}  \frac{dr}{2 \pi}  =[A+B \log{(k \epsilon)}]\left(\frac83+\frac{\sqrt{3}}{2 \epsilon}\right)
 \end{equation}
 $A$ and $B$ being constants depending on the integration interval. 
 Fixing $\epsilon$, and thus the black-hole spin, we can see that $N_{\rm cycles}$ depends on $\log k$, 
 and diverges when $k\to0$~\cite{Berti:2009bk}.

\begin{center}
\begin{figure}
       \includegraphics[width=6.cm,angle=-90]{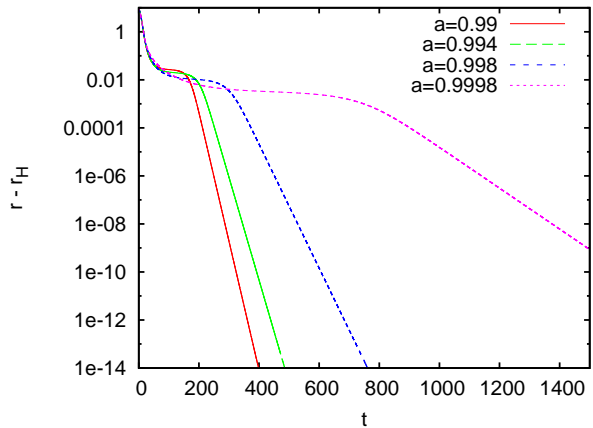} 
      \caption{The radial evolution with coordinate time, for geodesics having
$E=(E_{\max}+E_{\min})/2=2\epsilon$ and $L=b_{\rm ph}E (1-k)$ with $k=\mu=10^{-5}$, in Kerr spacetimes with
$a=0.99$, $0.994$, $0.998$ and $0.9998$. In particular, $r$ and $r_H$ are the Boyer-Lindquist coordinates of the particle and of the horizon.}
\label{fig:orbits}
 \end{figure}
\end{center}

In Sec.~\ref{comparison} we will study gravitational-wave emission numerically for 
geodesics having 
$E=(E_{\max}+E_{\min})/2=2\epsilon$, $L=b_{\rm ph}E (1-k)$ with $k=\mu=10^{-5}$, in Kerr spacetimes with 
$a=0.99$, $0.994$, $0.998$ and $0.9998$. In particular, we integrate these geodesics numerically (using a 4$^{\rm th}$ order Runge-Kutta integrator
with Richardson extrapolation). The Boyer-Lindquist radius a function of coordinate time is shown in Fig.~\ref{fig:orbits}. As can be seen, after an initial
plunge, the orbits show a plateau (corresponding to the particle orbiting at the light-ring) followed by the final plunge towards the horizon. The time spent at the light-ring
increases with $a$ [\textit{cf.} Eq.~\eqref{cycles}], and the horizon radius is
approached in an infinite coordinate time, in agreement with the analytical behavior pointed out by Ref.~\cite{MB}.
In order to estimate $N_{\rm cycles}$, we count the number of cycles of these numerically-integrated geodesics from $r=1.05 r_{\rm ph}$ to $r=(1.9 r_{\rm ph}+0.1 r_{\rm hor})/2$ ($r_{\rm ph}$ being the light-ring radius).
These values for  $N_{\rm cycles}$ can be obtained from  Eq.~\eqref{cycles} with 
$A\approx 0.620$, $B\approx0.01515$, to within 2.5\% ($\sim 0.17$ cycles).

\section{The gravitational-wave fluxes}\label{fluxes_sec}
\subsection{Analytics}
\label{analytics}
%
 The radiation emitted by point particles on circular geodesics in the Kerr geometry was studied 
semi-analytically by Chrzanowski and by Chrzanowski \& Misner, among others~\cite{Chrzanowski:1976jy,Chrzanowski:1974nr,breuer,kesden}. 
The original motivation for these investigations was the possible existence of 
synchrotron radiation in the Kerr geometry, a possibility that was ruled out by these studies. 
Performing a WKB analysis of the Teukolsky equation, Chrzanowski~\cite{Chrzanowski:1976jy} and Chrzanowski \& Misner~\cite{Chrzanowski:1974nr} 
concluded that the energy flux by a particle on the innermost stable circular orbit (ISCO) is {\it suppressed}
as the geometry approaches the extremal Kerr geometry. In particular,  they
found that the flux and therefore the total energy radiated by a particle 
with mass $\mu$ scales as
\be\label{E1scaling0}
E_1\sim (r-r_H) \mu^2\,,
\ee
where the orbital radius $r$ approaches the horizon radius $r_H$ in the extremal limit (because $r_{\rm ISCO}=r_H$ for $a=1$, cf. Ref.~\cite{ted_extremal_isco}).

This suppression of the energy flux in the extremal Kerr 
limit is common to other massless fields, like scalar or electromagnetic 
fields~\cite{Chrzanowski:1976jy,Chrzanowski:1974nr,breuer}. A possible interpretation 
for this behavior was put forward by Breuer~\cite{breuer}, who argues that in the $a\to 1$ limit,
a particle close to the ISCO approaches a principal null direction of the 
Kerr geometry, for which tidal and radiation effects are strongly suppressed~\cite{Penrose:1960eq,Stewart:1973vf}. 

Alternatively, Eq.~\eqref{E1scaling0} can be derived from dimensional arguments, which suggest that $d E/d\tau \sim \mu^2$ [$\tau$ being the particle's
proper time; see also Eq.~\eqref{dedtau}], with the factor $(r-r_H)$ coming from the gravitational redshift at the ISCO. In fact, for almost
extremal black holes the ISCO is very close to the horizon, and for near-horizon orbits one has $dt/d\tau \sim r_H/(r-r_H)$~\cite{MB}.
This second derivation suggests that the scaling \eqref{E1scaling0} should be valid also for the fluxes falling into
the black-hole horizon, and not
only for the fluxes at infinity studied by Refs.~\cite{Chrzanowski:1976jy,Chrzanowski:1974nr,breuer}. 
 In fact,  Kesden~\cite{kesden} has  recently revisited the problem of studying the gravitational emission by a particle on the ISCO of a Kerr spacetime. He used 
 the frequency-domain
 Teukolsky code GREMLIN~\cite{gremlin}, which can calculate both the fluxes at infinity and those falling into the horizon,
 and confirmed that the scaling~\eqref{E1scaling0} holds not only for the outgoing fluxes but also for the \textit{total} ones.

Because the ISCO approaches the light ring in an almost extremal Kerr spacetime (the ISCO, light-ring and horizon
actually coincide in the extremal case~\cite{ted_extremal_isco}), it should be possible to use the Eq.~\eqref{E1scaling0}
to infer the scaling of the gravitational radiation emitted by a photon at the light-ring. The same scaling will then hold also
for ultrarelativistic particles (i.e. ones with $E/\mu\gg1$) that orbit near the light-ring many times, 
such as those in which we are interested in this paper (cf. Sec~\ref{orbits}). 
This is because any null geodesic (in a generic geometry) can be obtained
as the limit of a series of timelike geodesics for which the particle's 4-momentum is kept finite and the particle's
rest-mass goes to zero. [More intuitively,
this amounts to saying that in General Relativity the (positive-energy) orbits of a neutrino, which has a small but non-zero mass, are very 
similar to those of a photon.]

Therefore, in order to take the null-geodesic limit of Eq.~\eqref{E1scaling0}, we first have to re-express it in terms of the 
energy $E$ of the (massive) particle at the ISCO. Because for such a particle $E\sim \mu$, we can recast
 Eq.~\eqref{E1scaling0} in the form
\be\label{E1scaling1}
E_1 \sim (r-r_H) E^2\,.
\ee
Because when $a\to 1$ the ISCO radius approaches the light-ring, this scaling must be also valid for a particle on
an unstable circular orbit (such orbits exist between the ISCO and the light-ring~\cite{bardeen}). The energy of the unstable circular orbits 
becomes positive near the light-ring (more specifically, it is positive for orbits with radius $r_{\rm ph}<r<r_{\rm mb}$, 
where $r_{\rm mb}$ is
the radius of the marginally bound circular orbit~\cite{bardeen}). We can then take a series of unstable circular orbits with constant positive energy
$E$ and with radius approaching the light-ring. The rest-mass of these orbits will clearly go to zero as $r\to r_{\rm ph}$, 
because the \textit{specific}
energy $E/\mu$ must diverge at the light-ring~\cite{bardeen}. Therefore, this series of timelike geodesics satisfies the conditions, 
outlined above, necessary to approach a null geodesics. This null geodesics will of course be a circular orbit at the light-ring with the same energy $E$ as the sequence
of unstable circular orbits that we have just mentioned, and its flux must too be described by the scaling~\eqref{E1scaling1}.

We stress that this result makes sense physically because in General Relativity it is 
the energy and not the rest-mass that gravitates. In fact, a more direct
derivation of Eq.~\eqref{E1scaling1} for a photon at the light-ring should be possible by solving the Teukolsky equations
using the stress-energy tensor for a photon, which is
\begin{equation}\label{eq:stress_en_ph}
{T}_{\rm ph}^{\mu\nu}(x) =\int {p}^\mu(\lambda) {p}^\nu(\lambda)\, \frac{\delta^{(4)}(x-z(\lambda))}{(-g)^{1/2}}\, d\lambda\,,
\end{equation}
where $z(\lambda)$ is the wordline and $p^\mu=dz^\mu/d\lambda$ is the 4-momentum.
(See for instance Ref.~\cite{deFelice:1990hu} for a derivation of the stress-energy tensor of the electromagnetic field
in the geometric-optics limit.) 
Clearly, ${T}_{\rm ph}^{\mu\nu}$ is non-zero despite the mass of the photon being zero, 
because the photon does have an energy, which curves the geometry. Of course, ${T}_{\rm ph}^{\mu\nu}$ agrees with
the stress-energy tensor of massive particle [cf. Eq. (19.3) of Ref.~\cite{poisson_rev}], when the mass is sent to zero keeping the 4-momentum finite.

Assuming then the validity of Eq.~\eqref{E1scaling1} for ultrarelativistic orbits and for photons as a working hypothesis, to be confirmed
numerically later in this paper, we can explore its consequences for the JS orbits considered in the previous section.
For those orbits, $E\sim r-r_H\sim \epsilon$ and we therefore expect
\be\label{E1scaling}
E_1\sim  \epsilon^3\,.
\ee
Also, we can now derive the behavior of the 
total radiated energy for high-energy plunges with near-critical impact parameter.
Because for these orbits the particle spends most of its time on almost circular orbits at the light ring, one can write
\be\label{totalfluxes}
E_{\rm rad}= \Delta E(\epsilon)\times N_{\rm cycles}\,,\quad
L_{\rm rad}= \Delta L(\epsilon)\times N_{\rm cycles}\,,
\ee
where $\Delta E$ and $\Delta L$ are the fluxes in a single orbit.
From a frequency-domain analysis, $\Delta E/\Delta L$ must equal the
light-ring frequency, $\Omega_{\rm ph}\approx 1/2 -(\sqrt{3}/2)\epsilon$, hence
\beq
\Delta E(\epsilon)&=&E_1(\epsilon) (1+e_2 \epsilon)\,,\label{scalingE}\\
\Delta L(\epsilon)&=&2 E_1(\epsilon) [1+(\sqrt{3} + e_2) \epsilon]\label{scalingL}\,.
\eeq
Here $e_2$ is an undetermined coefficient.

\subsection{Numerics}
\label{numerics}
\noindent 
The Teukolsky master equation describes scalar, vector and tensor field perturbations 
in the space-time of Kerr black holes~\cite{Teukolsky:1972my}. In Boyer-Lindquist coordinates, 
this equation takes the form
\begin{eqnarray}
\label{teuk0}
&&
-\left[\frac{(r^2 + a^2)^2 }{\Delta}-a^2\sin^2\theta\right]
         \partial_{tt}\Psi
-\frac{4 a r}{\Delta}
         \partial_{t\phi}\Psi \nonumber \\
&&- 2s\left[r-\frac{r^2-a^2}{\Delta}+ia\cos\theta\right]
         \partial_t\Psi\nonumber\\  
&&
+\,\Delta^{-s}\partial_r\left(\Delta^{s+1}\partial_r\Psi\right)
+\frac{1}{\sin\theta}\partial_\theta
\left(\sin\theta\partial_\theta\Psi\right)+\nonumber\\
&& \left[\frac{1}{\sin^2\theta}-\frac{a^2}{\Delta}\right] 
\partial_{\phi\phi}\Psi +\, 2s \left[\frac{a (r-1)}{\Delta} 
+ \frac{i \cos\theta}{\sin^2\theta}\right] \partial_\phi\Psi  \nonumber\\
&&- \left(s^2 \cot^2\theta - s \right) \Psi = -4\pi (r^2 + a^2 \cos^2 \theta)\, T ,
\end{eqnarray}
where $\Delta = r^2 - 2 r + a^2$ and $s$ is the ``spin weight'' of the field. 
The $s = -2$ version of this equation describes the evolution of the Weyl scalar 
$\psi_4 = \Psi / (r - i a \cos\theta)^4$ that characterizes the outgoing gravitational 
radiation. This is sufficient for the present work, because in order to determine the final spin
$a_f$ one only needs to account for the gravitational-wave radiation that leaves the binary system, while
 the fluxes emitted by the particle into the black-hole horizon (``ingoing fluxes'') cancel out when computing
the total angular momentum and energy and therefore do not affect the final spin parameter.\footnote{It is conceivably possible,
however, that the ingoing fluxes might overspin the black hole and possibly create a naked singularity \textit{before} the particle
is captured. We will comment on this effect in Sec.~\ref{comparison}.}

Computing the radiative energy and angular momentum loss is fairly straightforward from 
$\Psi$. More specifically, we use the following expressions for the radiated energy and angular momentum fluxes  
\begin{eqnarray}
\frac{{\rm d} E}{{\rm d} t}&=&\lim_{r\to\infty}\left\{\frac{1}{4
\pi r^6} \int_{\Omega} {\rm d}\Omega \left| \int_{-\infty}^{t}
\,{\rm d}\tilde{t}\,\Psi(\tilde{t},r,\theta,\varphi)\right|^{2}\right\} \nonumber\\
\frac{{\rm d} L_z}{{\rm d}t}&=&-\lim_{r\to\infty}\left\{\frac{1}{4
\pi r^6}\mathbf{Re}\left[\int_{\Omega} {\rm d}\Omega
\left(\partial_{\varphi}\,\int_{-\infty}^{t} \,
{\rm d}\tilde{t}\,\Psi(\tilde{t},r,\theta,\varphi)\right)\right.\right.\nonumber\\
&&\left.\left.\times\left(\int_{-\infty}^{t} \,{\rm d}t'\,
\int_{-\infty}^{t'} \,
{\rm d}\tilde{t}\,\bar{\Psi}(\tilde{t},r,\theta,\varphi)\right)\right]\right\}\;.
\end{eqnarray}
The variable $T$ on the right hand side of Eq.\ (\ref{teuk0}) is a point-particle source-term 
constructed from the energy-momentum tensor describing a particle moving in a Kerr spacetime.  
The particle energy-momentum tensor has the form 
\begin{eqnarray}
T_{\alpha\beta} &=& \mu \frac{u_\alpha u_\beta}{\Sigma \dot t\sin\theta}
\delta\left[r - r(t)\right]
\delta\left[\theta - \theta(t)\right]
\delta\left[\phi - \phi(t)\right]\;.
\end{eqnarray}
where $u_\alpha$ is the 4-velocity of the particle. It is noteworthy that $\dot t \equiv dt/d\tau$ 
appears in the denominator of this expression. As the particle approaches the horizon, $\dot t \to \infty$, 
which is just the well-known ``infinite redshift'' effect at the horizon of a black hole. Thus the particle 
source-term smoothly ``redshifts away'' as the particle approaches the horizon, therefore allowing the Teukolsky 
equation to gradually transition into its homogeneous form. This smoothly and naturally connects the gravitational 
radiation from the last few orbital cycles to the black hole's quasi-normal modes.     

To solve Eq.\ (\ref{teuk0}) numerically in time-domain we take the approach first introduced by Krivan 
\textit{et al.} in Ref.~\cite{Krivan1997}. Our code that solves the Teukolsky equation uses the same approach, 
therefore the contents of this section are largely a review of the work presented in the relevant literature. 

Our time-domain code uses the tortoise coordinate $r^*$ in the radial direction and azimuthal coordinate 
$\tilde{\phi}$. These coordinates are much better suited for performing numerical evolutions, as detailed 
in Ref.~\cite{Krivan1997}. They are related to the usual Boyer-Lindquist coordinates through the equations,  
\begin{eqnarray}
dr^* &=& \frac{r^2+a^2}{\Delta}dr 
\end{eqnarray}
and
\begin{eqnarray}
d\tilde{\phi} &=& d\phi + \frac{a}{\Delta}dr \; . 
\end{eqnarray}  
Following Ref.~\cite{Krivan1997}, we factor out the azimuthal dependence by using the $m$-mode decomposition,
\begin{eqnarray}
\Psi(t,r^*,\theta,\tilde{\phi}) &=& e^{im\tilde{\phi}} r^3 \Phi(t,r^*,\theta).
\end{eqnarray}
Defining
\begin{eqnarray}
\Pi &\equiv& \partial_t{\Phi} + b \, \partial_{r^*}\Phi \; , \\
b & \equiv &
\frac { {r}^{2}+{a}^{2}}
      { \Sigma} \; , 
\end{eqnarray}
where
\begin{eqnarray}
\Sigma^2 &\equiv &  (r^2+a^2)^2-a^2\,\Delta\,\sin^2\theta
\; 
\end{eqnarray} 
allows the Teukolsky equation to be rewritten in first-order form
\begin{eqnarray}
\partial_t \mbox{\boldmath{$u$}} + \mbox{\boldmath{$M$}} \partial_{r*}\mbox{\boldmath{$u$}} 
+ \mbox{\boldmath{$Lu$}} + \mbox{\boldmath{$Au$}} =  \mbox{\boldmath{$T$}} ,
\end{eqnarray}
where 
\begin{equation}
\mbox{\boldmath{$u$}}\equiv\{\Phi_R,\Phi_I,\Pi_R,\Pi_I\}
\end{equation}
is the solution vector. The subscripts $R$ and $I$ refer to the real and imaginary parts 
respectively (recall that the Teukolsky function $\Psi$ is a complex valued quantity). 
Explicit forms for the matrices {\boldmath{$M$}}, {\boldmath{$A$}} and {\boldmath{$L$}} 
can be easily found in the relevant literature~\cite{Krivan1997}. 

Lastly, an explicit time-evolution numerical scheme is developed for this first-order, 
linear PDE system using the two-step, 2nd-order Lax-Wendroff finite-difference method. 
Explicit details on this approach can be found in Ref.~\cite{Krivan1997}. Symmetries of 
the spheroidal harmonics are used to determine the angular boundary conditions: For even 
$|m|$ modes, we have $\partial_\theta\Phi =0$ at $\theta = 0,\pi$ while $\Phi =0$ at $\theta = 0,\pi$ 
for modes of odd $|m|$. We set $\Phi$ and $\Pi$ to zero on the inner and outer radial boundaries.

\begin{figure}
\centering
\includegraphics[width=6cm,angle=-90]{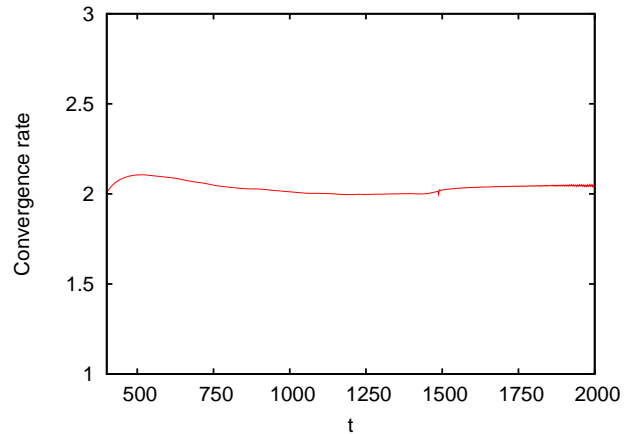}
\caption{Demonstration of our time-domain Teukolsky code's convergence rate, for $a=0.9998$ and $m=2$. \label{conv}}
\end{figure}

One major challenge in numerically solving Eq.\ (\ref{teuk0}) is representing a point-like 
particle source on a numerical grid. There are multiple approaches towards tackling this problem, 
such as representing the particle as a narrow Gaussian distribution~\cite{TDEMRICode} or taking a 
more efficient ``discrete Dirac-delta'' approach as presented in Ref.~\cite{TDEMRICodeMIT12}. 
One observation that we make specific to this work is that for near-extremal black holes in Boyer-Lindquist 
coordinates, the particle's orbit, the light-ring and the horizon are extremely close, therefore we
 modeled the point-particle to have a fixed width in the tortoise coordinate $r^*$ as opposed to $r$.
 This allows our code to resolve these distinct regions, because they are relatively widely separated in $r^*$. 
To test that our entire numerical scheme is working properly we performed extensive convergence tests on the data 
generated by our code, especially for the near-extremal Kerr hole cases. For these convergence tests, we kept the 
ratio of the particle's width to grid spacing constant \textit{i.e.} upon doubling grid-density we consistently halved the 
particle width. Sample convergence results for the $a = 0.9998$ and mode $m = 2$ case are presented in Fig.\ \ref{conv}.
 We observe a convergence rate that is extremely close to the expected 2nd-order convergence.  

One possible complication in our analysis comes from the fact that the convergence rate is actually slightly faster
than the theoretical 2nd-order convergence, as can be seen from Fig.\ \ref{conv}. This makes 
a Richardson extrapolation to further reduce discretization errors difficult to implement (because
one should account for the changes of the convergence rate along the trajectory in order to obtain an accurate result). Therefore,
we simply decide to decrease the grid spacing and particle's width until our numerical results change by no more than 0.5\%. 
The final grid spacing that we use in this work is $1/100$ in the $r^*$-coordinate and $0.02$ radians in the $\theta$-grid, and we use 
$1/500$ as the time-step. We also verify a posteriori that our numerical errors are less than 0.5\% by checking that the ratio 
$\Delta E/\Delta L$ coincides with the light-ring frequency to within 0.35\%.

Our code that implements this numerical scheme is a Fortran-code that is fully parallelized 
to execute efficiently on a computer cluster. The parallelization approach taken is the standard domain-decomposition approach 
(on the radial-coordinate numerical grid) with message-passing enabled using {\em OpenMPI}. 
Good scaling has been observed for several hundred processor-cores. In this current work, 
we made use of 200 processor-cores for computing each $m$-mode for every case that we studied. 
\subsection{Summing over $m$-modes}
\label{extrapolation}

\noindent As noted above, our evolution code evolves each $m$-mode separately. 
Therefore, we obtain the fluxes and radiated energy and angular momentum for each 
$m$-mode through distinct and separate numerical simulations. In order to compute 
the {\em total} radiated quantities, we would need to perform a sum over the results obtained 
from these different simulations. However, for higher $m$-modes it becomes increasingly difficult 
to perform accurate numerical computations mainly due to the requirement of significantly higher 
resolution (higher $m$-modes involve higher values of $\ell$ that require higher angular grid resolution 
for accurate representation; there is also a Courant condition that requires us to reduce the time-step
with higher angular resolution, thus making the computation even more demanding). Therefore, 
we use an alternative approach to estimate the radiated quantities from $m$-modes higher than 
$m=10$. For the case of circular and equatorial orbits, Finn \& Thorne~\cite{finn_thorne_00} show that  
\begin{equation}
{\dot E}_m=\frac{2(m+1)(m+2)(2m+1)!\, m^{2m+1}}{(m-1)[2^m\, m!\, (2m+1)!!]^2}\,\eta^2{\tilde \Omega}^{2+2m/3}\,{\dot{\cal E}}_{\infty\, m}
\label{series}
\end{equation}
where ${\tilde \Omega}$ in Eq.\ (\ref{series}) is the Keplerian angular frequency 
of the orbit i.e. ${\tilde \Omega}^{2/3} = 1/r$. 

Because the gravitational-wave fluxes for JS orbits are produced, for the most part, during the quasi-circular equatorial
cycles at the light-ring, our numerically calculated fluxes should satisfy this scaling, and in particular 
they should be in a geometric progression for large values of $m$, because  Eq.~\ref{series} implies 
${\dot E}_{m+1}/{\dot E}_m\longrightarrow {\rm constant}$ for $m\gg 1$. We have indeed verified that this is the case, \textit{i.e.}
we have checked that ${\dot E}_{m+1}/{\dot E}_m$ and ${\dot L}_{m+1}/{\dot L}_m$ are constant to within 0.5\% already for $m\gtrsim 9$.

This behavior allows us to reconstruct the total fluxes in the following way: 
(1) We use our Teukolsky evolution code described above 
to compute the fluxes and the radiated quantities for $m$-modes up to $m=10$; (2) We use  
the fluxes calculated for $m=9$ and $m=10$ to estimate the ratio entering the geometric progression 
described above and thus obtain an estimate of the contribution from the higher $m$-modes; 
(3) We finally add in this estimate to the explicitly computed values in step (1) and thus obtain the 
{\em total} radiated amount for both the energy and the angular momentum.  

We stress that while the fluxes summed up to $m=10$ are accurate at least to within 0.5\% (\textit{cf.} discussion in the previous section),
the procedure that we have just described introduces larger errors into the total fluxes. For the 
orbits that we consider in this paper, the asymptotic ratios ${\dot E}_{m+1}/{\dot E}_m$ and ${\dot L}_{m+1}/{\dot L}_m$ grow from 0.928 
for $a=0.99$ to 0.94 for $a=0.9998$. 
Since the sum of a geometric series is proportional to $1/(1-r)$, $r$ being the asymptotic ratio, a 0.5\% error in estimating $r$ would lead 
to a $\sim$8\% error in the sum of the fluxes with $m>10$. This propagates into a significant error in the total fluxes, because for relativistic 
plunging orbits around almost-extremal Kerr black holes such as those that we are considering in this paper, the contribution of the large multipole moments
to the total fluxes decays slowly with $m$. In particular, because the contribution of the fluxes with $m>10$  grows from 1.25 times 
the contribution of the fluxes with $m=0-10$ for $a=0.99$, up to 1.7 times for  $a=0.9998$, 
we estimate the error of our total fluxes to be $\lesssim 5$\%.

\section{Comparing the analytical and numerical gravitational-wave fluxes}
\label{comparison}

In our previous \textit{Letter}~\cite{letter}, we considered black holes with $a=0.99$, $0.992$, $0.994$, $0.996$, $0.998$, $0.999$ and $0.9998$,
and geodesics having $E=(E_{\max}+E_{\min})/2=2\epsilon$, $L=b_{\rm ph}E (1-k)$ with $k=10^{-5}$, and $\mu=0.001$.
Using the time-domain Teukolsky code that we described in Sec.~\ref{numerics}, we verified that the theoretical 
scaling given by Eqs.~\eqref{E1scaling}-\eqref{scalingL} works well for $a<0.999$, but predicts fluxes that
are too large for $a=0.999$ and $a=0.9998$. More specifically,  for $a<0.999$ the deviations of the numerical results from
the theoretical scaling are about $1-3\%$ and therefore comparable to the errors discussed in Sec.~\ref{numerics}. However,
for $a=0.999$ and  $a=0.9998$ the fluxes predicted by Eqs.~\eqref{E1scaling}-\eqref{scalingL} are respectively $12\%$ and 84\% larger than
the numerical ones. 

Because our \textit{Letter}~\cite{letter} was only concerned about whether the gravitational-wave fluxes could prevent
naked singularities from forming, this discrepancy did not affect our conclusions. In particular, it reinforced 
our finding that there are orbits giving rise to naked singularities \textit{even} 
when radiation reaction is taken into account. In this section, however, we will investigate the origin of this discrepancy
between our theoretical scaling and our numerical results.

First, let us note that a crucial assumption in the derivation of the scaling~\eqref{E1scaling}-\eqref{scalingL} is that
the orbits be relativistic. In fact, only under that assumption we can expect the fluxes to depend only on the energy $E$ (and therefore
on $\epsilon$) and not on the rest mass $\mu$. However, because the orbits considered in Ref.~\cite{letter} 
have energy $E=(E_{\max}+E_{\min})/2=2\epsilon$
and mass $\mu=10^{-3}$, the ratio $E/\mu$ is $\sim22$ for $a=0.999$ and $10$ for $a=0.9998$. Therefore, a possible explanation for the 
disagreement between our analytical and numerical results is that the orbits that we considered are simply not relativistic enough.
In order to test this hypothesis, we have tried producing orbits that have exactly the same energy and angular momentum
as those considered in Ref.~\cite{letter}, but which have different rest mass. More specifically, we have 
considered black holes with $a=0.99$, $0.992$, $0.994$, $0.996$, $0.998$, $0.999$ and $0.9998$,
and geodesics having $E=(E_{\max}+E_{\min})/2=2\epsilon$, $L=b_{\rm ph}E (1-k)$ with $k=\mu=10^{-5}$. If the orbits
with $\mu=10^{-3}$ were already sufficiently relativistic to satisfy the scaling~\eqref{E1scaling}-\eqref{scalingL}, then their
fluxes should coincide, to within the numerical errors, with those of the orbits with $\mu=10^{-5}$, since the two set of orbits have exactly
the same energies and angular momenta.

\begin{figure}
\centering
\includegraphics[width=6.3cm,angle=-90]{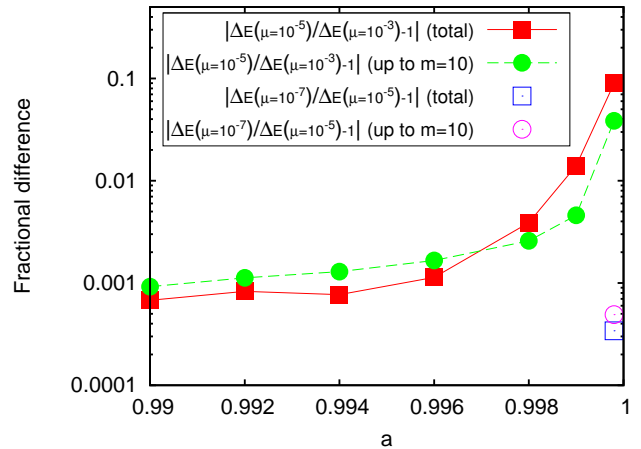}
\caption{\label{test} Fractional difference between 
the energy fluxes emitted in a single orbit at the light ring for $\mu=10^{-3}$ and $\mu=10^{-5}$ (filled symbols), 
and for $\mu=10^{-7}$ and $\mu=10^{-5}$ (empty symbols). 
The fluxes are obtained by summing over all multipole moments (``total'' fluxes, represented by squares) 
using the prescription outlined in Sec.~\ref{extrapolation}, or by summing only up  to $m=10$ (fluxes ``up to $m=10$'', represented by circles).}
\end{figure}

In Fig.~\ref{test}, we therefore show the fractional difference between the energy fluxes (in a single orbit at the light ring) for the two sets of orbits. In particular, 
we have calculated this difference both for the total fluxes (obtained by summing over all multipole moments $m$ as described in Sec.~\ref{extrapolation})
and for the ``partial'' fluxes obtained   by summing up to $m=10$. As can be seen the difference is less than 0.5\% for $a<0.999$, is about $1$\%
for $a=0.999$ and grows to $\sim$10\% for $a=0.9998$, thus suggesting that at least in this last case the orbit with $\mu=10^{-3}$ is not sufficiently
relativistic. As a further confirmation, we have produced numerical fluxes for a particle moving in a Kerr spacetime with $a=0.9998$, and having
the same energy and angular momentum as the previous orbits but rest mass $\mu=10^{-7}$. As can be seen from Fig.~\ref{test}, this orbit gives fluxes
that agree with those produced for the $\mu=10^{-5}$ orbit to within less than 0.1\%. 

This test confirms that orbits with $\mu=10^{-5}$ are sufficiently relativistic for the scaling~\eqref{E1scaling}-\eqref{scalingL} to hold. 
However, although using these orbits improves the agreement with the expected scaling, the deviations remain as large as 5\% for $a=0.999$
and $70$\% for $a=0.9998$. In fact, a careful analysis of the convergence of all our numerical fluxes with respect to the grid spacing and 
particle's width revealed that the modes with $m\gtrsim 8$ did not completely converge in the case of the orbits around black holes with $a=0.999$
and $a=0.9998$ considered in our previous work~\cite{letter}. This convergence error was then amplified by the procedure that we use to
reconstruct the contribution of the large-$m$ modes. (As we discussed in Sec.~\ref{extrapolation}, this procedure is very sensitive to
 the numerical results for $m=9$ and $m=10$). 

 The resolution that we used was instead sufficient in the case of the other multipole
moments for  $a=0.999$ and $a=0.9998$, and for all the multipole moments at lower spins. 
Indeed, it is not surprising that this problem affected only the high-$m$
modes for spins very close to 1. When $a\sim 1$, the frequency of the light-ring becomes very close to that of the horizon, and so does the radius 
(in Boyer-Lindquist coordinates). It is therefore necessary, as already mentioned in Sec. \ref{numerics}, to use a very high resolution
in the near-horizon region to calculate the fluxes accurately for $m\gg 1$ (because large $m$'s correspond to small lengthscales).

In order to amend this problem, we focused on just four orbits -- namely orbits around black holes with $a=0.99$, $0.994$, $0.998$ and $0.9998$,
and having $E=(E_{\max}+E_{\min})/2=2\epsilon$, $L=b_{\rm ph}E (1-k)$ with $k=\mu=10^{-5}$ -- but calculated the fluxes with very 
high grid resolution and small particle's width. More specifically, as explained in Sec. \ref{numerics}, we estimated the error connected to 
the finite grid resolution and particle's width to be less than 0.5\%, and the error due to the reconstruction of the large-$m$ modes to be less 
than 5\%.

As discussed in Sec. \ref{orbits}, the number of cycles described by these orbits at the light ring
is reproduced by Eq.~\eqref{cycles} with $A\approx 0.620$ and $B\approx0.01515$ to within 2.5\%.
Assuming then $E_1= e_1 \epsilon^n$, we fit the numerical energy and angular-momentum 
fluxes with Eqs.~\eqref{totalfluxes}-\eqref{scalingL}, obtaining $n\approx 2.95$.
Because this value is very close to the theoretical value $n=3$ [\textit{cf.} Eq.~\eqref{E1scaling}], we then assume 
$n=3$ and fit the data with only two free parameters, $e_1$ and $e_2$,
obtaining $e_1=233.72$ and $e_2=-5.83$.\footnote{This value of $e_1$, which represents the overall normalization of the fluxes emitted in
a single orbit, is significantly different from that reported in Ref.~\cite{letter}. This is due to the different ways of calculating the 
number of orbits $N_{\rm cycles}$ at the light ring that we used in this paper and in Ref.~\cite{letter}. In Ref.~\cite{letter}, we defined $N_{\rm cycles}$ as the number of cycles between $r=1.05 r_{\rm ph}$ and $r=(r_{\rm ph}+r_{\rm hor})/2$, while here we decided to bracket the light ring more 
closely and considered the number of cycles between $r=1.05 r_{\rm ph}$ and $r=(1.9 r_{\rm ph}+0.1 r_{\rm hor})/2$. These different definitions are also the reason why the values of $A$ and $B$ that we report in this paper are significantly different from those of Ref.~\cite{letter}.} 
With these values, Eqs.~\eqref{E1scaling}-\eqref{scalingL} 
reproduce the numerical data with residuals $\lesssim 4$\%. As discussed in Secs.~\ref{numerics} and \ref{extrapolation}, these
residuals are comparable to the errors affecting our total fluxes, thus confirming our expected scaling.

Utilizing Eq.~\eqref{radcorrections}, it is now straightforward to determine the consequences of this scaling for the final spin $a_f$.
From Eqs.~\eqref{totalfluxes}--\eqref{scalingL}, we obtain
\begin{equation}
L_{\rm rad}-2E_{\rm rad}= 2 \sqrt{3}\epsilon E_1 N_{\rm cycles}\,,
\end{equation}
which implies in particular $L_{\rm rad}-2E_{\rm rad}>0$, \textit{i.e.} the fluxes tend to decrease the final spin.
Moreover, using expression~\eqref{cycles} for $N_{\rm cycles}$ and  expression \eqref{E1scaling} for $E_1$,
we obtain that $L_{\rm rad}-2E_{\rm rad}\sim  \epsilon^3\log(k\epsilon)$. Comparing this scaling with the ``overspinning''
term $8 \epsilon^2 (1-x) x y$ in Eq.~\eqref{radcorrections}, it is clear that if $k$ is kept constant and
$\epsilon$ is sent to 0 (\textit{i.e.,} if one considers initial spins larger than some critical value $a_{\rm crit}$),
the effect of the fluxes eventually becomes subdominant relative to the ``overspinning'' term.
In other words, the fluxes tend to decrease the final spin $a_f$, but their effect is not sufficient to 
prevent the formation of naked singularities if the initial spin is sufficiently close to 1. Our numerical results confirm
this picture, as can be seen from Table~\ref{tab:jfin}, where  
we show the initial spin $a$, the final spin $a_f^{\rm JS}$ as computed by JS (\textit{i.e.,} without including the effect of the fluxes),
and the final spin $a_f$, taking into account radiation reaction. As can be seen, for the orbits that we consider 
$a_f>1$ already for $a=0.9998$, and therefore the critical spin mentioned above is $a_{\rm crit}\approx 0.9998$ 
(corresponding to $\epsilon\approx0.01$~\cite{letter}).

\begin{table}
\begin{tabular}{rcccc}\hline\hline
$a$             & 0.99   &   0.994  &   0.998  & 0.9998 \\

$a_f$           & 0.886  &     0.963 &   0.997  & 1.00004 \\

$a_f^{JS}$       & 1.0043 &     1.0026 &  1.0009 & 1.00009 \\

\hline \hline
\end{tabular}
\caption{Initial and final black-hole spin after absorbing a relativistic particle 
with energy $E=\sqrt{2(1-a)}$ and angular momentum $L=b_{\rm ph}E (1-10^{-5})$, 
neglecting conservative self-force effects, but not radiation reaction. 
We also show the final spin without radiation reaction ($a_f^{JS}$) predicted by JS.}
\label{tab:jfin}
\end{table}

Finally, let us further comment on the effect of the ingoing fluxes. As already mentioned in Sec.~\ref{numerics}, the ingoing fluxes
do not affect the final spin, which only depends on the energy and angular momentum that leave the binary system. However, the 
energy and angular momentum fluxes falling into the horizon might in principle spin the black hole  up and even create a naked
singularity \textit{before} the particle is captured. 

Because our code only calculates the outgoing fluxes, we cannot test this conjecture explicitly. However,
we can notice that the analytic derivation of our scaling for the fluxes 
[Eqs. \eqref{E1scaling}--\eqref{scalingL}]
applies \textit{both} to the outgoing \textit{and} ingoing fluxes, because nowhere in the derivation we do make use 
of the fluxes being ingoing or outgoing. (Also, as already stressed, Ref.~\cite{kesden} found that the scaling \eqref{E1scaling0} from which we start holds
not only for the outgoing fluxes but also for the total ones.) Because we have validated our scaling [Eqs. \eqref{E1scaling}--\eqref{scalingL}]  
by comparing it to numerical
results for the outgoing fluxes, we can now use it to assess the effect of the ingoing ones. In particular, 
the spin of the black hole 
\textit{before} the particle is captured is given by
\begin{multline}
a^\prime=
1-2\epsilon^2+L_{\rm rad,in}-2E_{\rm rad,in}=\\1-2\epsilon^2+2 \sqrt{3}\epsilon E^{\rm in}_1 N_{\rm cycles}\,,
\label{intermediate_spin}
\end{multline}
where we have used
Eqs.~\eqref{totalfluxes}--\eqref{scalingL}
to estimate the ingoing fluxes $E_{\rm rad,in}$ and $L_{\rm rad,in}$. Using expressions 
\eqref{E1scaling} for $E^{\rm in}_1$ and \eqref{cycles} for $N_{\rm cycles}$,
we obtain $L_{\rm rad,in}-2E_{\rm rad,in}\sim  \epsilon^3\log(k\epsilon)$. As for the outgoing fluxes, this scaling
shows that for fixed $k$ the ingoing fluxes are negligible with respect to
the term $2\epsilon^2$ in  Eq.~\eqref{intermediate_spin}, if the initial spin is sufficiently large. In other words,
if the initial spin is larger than some critical value $a^\prime_{\rm crit}$, no naked singularity can form before the particle is captured.
However, if $a<a^\prime_{\rm crit}$, the effect of the ingoing fluxes is
dominant over the term quadratic in $\epsilon$. The exact value of $a^\prime_{\rm crit}$ will depend on the normalization
of the ingoing fluxes $E_{\rm rad,in}$ and $L_{\rm rad,in}$. However, since the ingoing fluxes are expected to be comparable to
the outgoing ones -- because the fluxes for JS orbits are produced when the 
particle sits at the light-ring, which roughly corresponds to the maximum of
the effective potential for gravitational waves -- $a^\prime_{\rm crit}$ 
should be similar to the critical spin $a_{\rm crit}\approx0.9998$ (corresponding to
$\epsilon\approx0.01$~\cite{letter}) 
relevant for the outgoing fluxes.

Even more worrisome, the ingoing energy flux $E^{\rm in}_1$ must be positive. This is because the ingoing energy flux
can be negative only in the superradiant regime, which would require that the particle's orbital frequency be smaller 
than the horizon frequency. Since $\Omega_{\rm ph}\approx1/2-(\sqrt{3}/2)\epsilon$ is 
larger than the horizon's frequency $\Omega_{\rm hor}\approx 1/2-\epsilon$, it must be $E^{\rm in}_1>0$, which implies
that the intermediate spin \eqref{intermediate_spin} is larger than the initial spin. 
In particular, for $a<a^\prime_{\rm crit}$,
the ingoing fluxes would overspin the black hole past the extremal limit and create a naked singularity before the particle is captured. 
We will see in the next section how the conservative self-force provides a mechanism which has the correct order of magnitude and (possibly)
the right sign to prevent both the formation of naked singularities due to the ingoing fluxes (for $a<a^\prime_{\rm crit}$),
and the formation of naked singularities due to the particle's capture (for $a>a^\prime_{\rm crit}$).

\subsection{Suppressing dissipative effects with a ring of particles}
\label{ring_section}

The above analysis shows that dissipative effects are not sufficiently
strong to invalidate JS analysis. In fact, there is another simple argument indicating that dissipative effects can
{\it almost} be neglected. Ours and JS's analysis considered a test particle of energy $E$, rest-mass $\mu$ and angular momentum $L$.
Take now instead $N$ particles each of energy $E/N$, rest-mass $\mu/N$ and angular momentum $L/N$.
Throw these $N$ particles separated by $2\pi/N$ radians along the equator. JS calculation proceeds in exactly the same way, and the same results will be obtained. However, radiation effects will be suppressed. In fact, when $N\to \infty$, radiation
should be suppressed completely. The reason is that the ring can be thought of as a sum of point particles, and the fluxes result from the
interference between the gravitational waveforms from all the particles. Generically, this interference {\it always reduces} the energy output
relatively to a single particle \cite{Haugan:1982fb,Shapiro:1982,Oohara:1983gq,Kojima:1984cj}. 

In fact, for rings with angular momentum close to the critical one, most of the radiation is emitted in a 
quasi-circular orbit, as we have argued before. Because the contribution from particles in circular orbits 
has terms of the form $\delta (\omega-m\Omega)$ and because $m=0$ for a ring, then the radiation 
in the circular regime is actually totally suppressed. This has been verified numerically by a number of authors.
For trajectories plunging into a Schwarzschild black hole Oohara and Nakamura found 
that radiation is suppressed as the angular momentum of the ring is increased \cite{Oohara:1983gq}, 
while Kojima and Nakamura found similar results for plunging rings into rotating black holes \cite{Kojima:1984cj}. 

Thus, the construction of a ring of JS-like particles is able to suppress radiation, while
violating the Cosmic Censorship Conjecture, just like in the original JS analysis.
This is our final argument showing why dissipative effects cannot prevent naked
singularities from forming in this scenario, and that instead conservative effects must be taken into account.
\section{The conservative self-force}
\label{conservative_sf_section}

In the previous section, we have shown that the radiation reaction can
prevent the formation of naked singularities only for \textit{some} of the JS orbits.
More specifically, the final spin $a_f$, which includes the effect of the radiated energy and angular momentum,
is smaller than $1$ only for orbits with impact parameter $b$ extremely close to
the light-ring's impact parameter $b_{\rm ph}$, and the difference $|b-b_{\rm ph}|$ needed to ensure $a_f<1$ becomes smaller and smaller
as the initial spin $a$ approaches $1$, \textit{i.e.} the radiation reaction becomes less and less effective near the extremal limit.

Also, we have shown that in the cases in which $a_f<1$, the gravitational-wave fluxes that enter the horizon before the particle
is captured tend to spin the black hole  up past the extremal limit, forming a naked singularity. While the capture of the particle
would then offset this spin-up and produce a final spin  $a_f<1$, it is unclear whether the capture would happen at all, 
because the black-hole horizon would have disappeared and the particle may simply be scattered by the naked singularity. Also, even if the
particle were captured, a naked singularity would exist for a finite amount of time, and the Cosmic Censorship Conjecture would still be violated.

We will now use the results of the previous section to 
estimate another effect, the so-called conservative self-force~\cite{poisson_rev,MST,QW,gralla,pound}, and show that
its order of magnitude is sufficient to solve both these problems and prevent the formation of naked singularities for
\textit{all} of the JS orbits, provided that it carries a certain sign. Because our approach can only estimate
the order of magnitude of the conservative self-force, but not its sign, our result can be viewed as putting forward
a simple test of the  Cosmic Censorship Conjecture using self-force codes.
Unfortunately, to date none of the existing self-force codes~\cite{SFcodes} can handle orbits that, like the JS orbits, are relativistic and
around almost-extremal Kerr black holes. As we will show, however, the existing calculations for non-relativistic 
orbits around non-spinning black holes seem to hint at a conservative self-force sign consistent with no formation of naked singularities.
While more general self-force codes than those available today will be needed to say the last word on whether the JS orbits might
create naked singularities, this result seems to hint at the conservative self-force playing a crucial role in enforcing the
 Cosmic Censorship Conjecture.

In order to introduce the concept of conservative self-force, let us consider a black hole  
with gravitational radius $R_g = 2 G \mu/c^2$, moving in a curved background spacetime with ${\cal L}\gg R_g$.\footnote{This discussion 
is completely general because the motion
of a black hole  is the same as that of a particle with the same mass $\mu$, at leading and next-to-leading order in $R_g/{\cal L}$~\cite{MST}.}
In order to study the motion of this black hole  in a completely rigorous way, one would need
 to set up a proper initial value formulation, but a reasonable
alternative for practical purposes is to use a matched asymptotic expansion~\cite{MST,gralla,pound}.
In particular, near the black hole  (\textit{i.e.} at distances $r$ to the black hole  smaller than 
some limiting radius  $r_i\ll {\cal L}$), the metric can be written as 
\begin{equation}
g_{\rm internal} = g_{\rm BH} +H_1(r/{\cal L})+H_2(r/{\cal L})^2+...\,,
\end{equation}
 where $g_{\rm BH}$ is the metric of an isolated black hole  and 
$H_1(r/{\cal L})$, $H_2(r/{\cal L})^2$ are corrections due to the presence of the ``external'' background.

Far from the black hole  (\textit{i.e.} for $r>r_e$, $r_e$ being a suitable radius $\gg R_g$), the geometry is that of 
the background spacetime plus perturbations due to the black hole's presence, and the metric can therefore be written 
as 
\begin{equation}
g_{\rm external} = 
g_{\rm background} +h_1 (R_g/{\cal L})+h_2 (R_g/{\cal L})^2+...\,,
\end{equation} 
$h_1$ and $h_2$ being functions (of time and position) representing the perturbations produced by the black hole.

Because $R_g \ll {\cal L}$, there exists a region $r_e<r<r_i$ where both pictures are  
valid and the two metrics can be matched. Doing so, one obtains  
that the black-hole  equations of motion are~\cite{MST,gralla,pound,QW}
\begin{equation} \label{sf}
u^\mu \nabla_\mu u^\nu = f^\nu_{\rm cons}+f^\nu_{\rm diss}+{\cal O}(R_g/{\cal L})^2\,,
\end{equation} 
where $\nabla$ is the Levi-Civita connection of the background spacetime. The terms $f^\nu_{\rm cons}$ 
and $f^\nu_{\rm diss}$ are ${\cal O}(R_g/{\cal L})$, and are dubbed the conservative and dissipative self-force. 
Remarkably, Eq.~\eqref{sf} turns out to be 
the geodesic equation of a particle in a ``perturbed'' metric $\tilde{g}=g+h^R$, where $h^R$ is a smooth tensor field
of order ${\cal O}(R_g/{\cal L})$:
\begin{equation}  \label{sf2}
\tilde{u}^\mu \tilde{\nabla}_\mu \tilde{u}^\nu = 0\,.
\end{equation} 
In this equation, the Levi-Civita connection $\tilde{\nabla}$ and the 4-velocity $\tilde{u}^\mu$ are
 defined with respect to the ``perturbed'' metric $\tilde{g}=g+h^R$, and $h^R$ can be interpreted as the (regularized) metric
perturbation produced by the black hole  itself.

The dissipative self-force can be shown~\cite{sf_diss} to be equivalent to the effect
of the energy and angular-momentum fluxes on the particle's trajectory, which we  considered in the previous sections. 
We can therefore use the results of Sec.~\ref{comparison}
to say something about the scaling of the self-force. Considering for instance the energy lost in gravitational waves, 
from the definition of the particle's energy $E=-p_t$ and from Eq.~\eqref{sf} one obtains
\be\label{dedtau}
dE/d\tau= - \mu f^{\rm diss}_t ={\cal O}(R_g/{\cal L})^2\,.
\ee 
Assuming now that the background spacetime is a black hole  with mass $M\sim {\cal L} \gg R_g$, and specializing to orbits near the horizon, 
from the geodesics equation one immediately obtains 
$dt/d\tau \sim r_H/(r-r_H)$~\cite{MB}, which implies
\be\label{dedt_sf}
dE/dt\sim (r-r_H) {\cal O}(R_g/{\cal L})^2\,.
\ee 
 Because in the extremal Kerr geometry the ISCO, the marginally bound circular orbit and the light ring coincide 
with the horizon~\cite{ted_extremal_isco}, 
this equation will be valid for the circular orbits between the ISCO and the light-ring in the almost-extremal case. Also, because 
for these orbits the energy $E$ is proportional to the mass $\mu$, we can think of $R_g$ as being proportional to the energy $E$
in Eq.~\eqref{dedt_sf} (this corresponds to replacing $\mu$ with $E$ when going from Eq.~\eqref{E1scaling0} to Eq.~\eqref{E1scaling1} in 
Sec.~\ref{analytics}). 
We can then
follow the same procedure that we used in Sec.~\ref{analytics}, i.e.
we can consider a sequence of unstable circular orbits between the marginally bound orbit and the light-ring, with constant
positive energy (and therefore with rest-mass going to zero as the light-ring is approached). Taking the limit of
Eq.~\eqref{dedt_sf} along this sequence of orbits shows that Eq.~\eqref{dedt_sf} should also be valid for a photon at the light-ring, provided
that $R_g$ is interpreted as proportional to the photon \textit{energy}.

As already stressed, the numerical results of Sec.~\ref{comparison} support this interpretation of Eq.~\eqref{dedt_sf}. In fact, 
for the JS orbits $R_g$ should scale like the particle energy $E\sim \epsilon$, and because 
$r_{\rm ph}-r_H\sim {\cal O}(\epsilon)$ we have that Eq.~\eqref{dedt_sf} gives $dE/dt\sim {\cal O}(\epsilon)^3$, which is indeed
the scaling that we found numerically in Sec.~\ref{comparison}. It therefore seems that
for a black hole  with $E\gg \mu$,
the size entering the matched asymptotic expansion analysis that we sketched above
 is $R_g = 2 G E/c^2$ and \textit{not} $R_g = 2 G \mu/c^2$. 
This is hardly surprising, since the size associated with a
black hole  or particle moving at relativistic speeds is given 
by its energy and not by its mass, simply because in General Relativity energy gravitates.

Further evidence comes from the so-called Aichelburg-Sexl metric, which represents a
Schwarzschild black hole  as seen by an observer moving at nearly the speed of light. (More precisely, 
the Aichelburg-Sexl metric can be obtained by boosting the Schwarzschild metric to the speed of light, keeping the total energy fixed.)
As one would expect from physical intuition, this metric depends on the total energy $E$ and not on the rest-mass~\cite{Aichelburg:1970dh},
and in particular this boosted black hole 
absorbs particles within a distance $\sim E$ from it.

To make the argument more rigorous, one may attempt to 
 set up a matched asymptotic expansion for a \textit{photon}
in a generic curved background. Because the stress-energy tensor of a photon [Eq.~\eqref{eq:stress_en_ph}] depends on its energy
and not on its rest-mass (which is of course zero),  and because the 
metric near a photon is presumably given by the Aichelburg-Sexl metric, the only size
entering the matched asymptotic expansion and therefore the self-force will be $R_g = 2 G E/c^2$.
Developing such a formalism goes beyond the scope of this paper, but the picture that we described above 
is physically clear, and it is remarkable that 
we were able to test it with the numerical results presented in Sec.~\ref{comparison}.

Based on this argument, the size of
a black hole  or particle moving on a JS orbit is given by  $E\sim \epsilon$, 
which is sufficient to conclude that the 
conservative self-force affects JS's analysis.
 This is easy to see from Eq.~\eqref{sf2} 
(although the same result can be obtained from Eq.~\eqref{sf}: see Ref.~\cite{barack_sago}): 
because the regularized metric ``perturbation'' $h^R$ is ${\cal O}(R_g/{\cal L})={\cal O}(\epsilon)$,
the effective potential for the radial motion differs from the ``geodetic'' one by   
${\cal O}(R_g/{\cal L})={\cal O}(\epsilon)$~\cite{barack_sago,Sago:2008id}.
In particular, the light ring's impact parameter $b_{\rm ph}$ becomes $b_{\rm ph}+\delta b_{\rm ph}$, with $\delta b_{\rm ph} ={\cal O}(R_g/{\cal L})={\cal O}(\epsilon)$. 
Because the JS orbits have $b_{\rm ph}-b={\cal O}(R_g/{\cal L})={\cal O}(\epsilon)$, 
the conservative self-force may prevent them from plunging into the horizon. This effect is intuitive: if the particle's 
size is $\sim \epsilon$, finite-size effects are important for impact parameters $b=b_{\rm ph}+{\cal O}(\epsilon)$ (\textit{i.e.} the impact 
parameter is so close to the light ring's impact parameter, which discriminates between plunging and scattering orbits, that finite-size
effects must be taken into account).

Clearly, what remains to be determined in this analysis is the coefficient of the impact parameter change $\delta b_{\rm ph}$ 
produced by the conservative self-force. In particular, the sign of  $\delta b_{\rm ph}$  is crucial, because if $\delta b_{\rm ph}>0$
the light ring's impact parameter would increase, actually making it easier for the JS orbits to plunge into the black hole  and produce
a naked singularity. If instead $\delta b_{\rm ph}<0$,  the light ring's impact parameter and therefore the black-hole  photon cross section would shrink, 
thus making it harder for 
the JS orbits to hit the black hole.

As we have already mentioned, a calculation of $\delta b_{\rm ph}$ is not 
feasible with present self-force codes~\cite{SFcodes}, which can only handle non-relativistic orbits around
non-spinning black holes. Nevertheless, we can try to use the existing results for these orbits to guess how the impact
parameter of the light ring might change under the effect of the conservative self-force. In particular,
Ref.~\cite{barack_sago} showed that the ISCO frequency in a Schwarzschild black hole  increases due to the conservative self-force.
If the same behavior applies to relativistic orbits in almost extremal Kerr black holes, \textit{i.e.}~if  $\Omega_{\rm ph}$ for these
spacetimes increases under the effect of the conservative self-force, then $b_{\rm ph}$ should decrease (\textit{i.e.,} $\delta b_{\rm ph}<0$),
because for circular photon orbits one has
$b_{\rm ph}= 1/\Omega_{\rm ph}$. (This follows from the fact that the photon 4-momentum is a null vector:  from $p^\mu p_\mu=0$ and 
from $p_\mu=-E \delta_\mu^t+L\delta^\phi_\mu$ and $p^\mu= p^t (\delta_t^\mu+\Omega \delta_\phi^\mu)$, one immediately
obtains $b=L/E=1/\Omega$.) This would imply, as already mentioned,
that the black-hole photon cross section would shrink under the effect of the conservative self-force, possibly preventing the JS orbits 
from being captured and
 naked singularities from being formed.

 Clearly, explicit self-force calculations for relativistic orbits around almost extremal Kerr black holes will
be needed to confirm this conjecture and determine the exact numerical value of $\delta b_{\rm ph}$ (in order to go beyond
the order of magnitude result $\delta b_{\rm ph} ={\cal O}(\epsilon)$ derived above). It is of course very well possible
that such explicit calculations will find that $b_{\rm ph}$ increases rather than decreases for almost extremal black holes, or that
its decrease is too small to prevent the JS orbits from being captured. Hints at a possible change of sign of $\delta b_{\rm ph}$
when going from $a=0$ to $a\sim 1$ come for instance from Ref.~\cite{WB}, in which Warburton and Barack found that 
the ISCO frequency for a \textit{scalar} particle \textit{increases} under the effect of the (scalar) self-force when $|a|\lesssim 0.9$, but
\textit{decreases} for $a\gtrsim 0.9$. A similar result was found by Refs.~\cite{eob2}, 
who calculated the ISCO shift due to the conservative
self-force using an effective one-body model for spinning black-holes~\cite{eob1,eob2}, 
calibrated in the $a=0$ case with the results of Ref.~\cite{barack_sago},
and found that the ISCO frequency decreases at high spins.
These results, albeit still inconclusive, highlight even more compellingly the need for a rigorous calculation
of the \textit{gravitational} self-force for \textit{ultrarelativistic} particles (or photons) moving in an \textit{almost extremal} Kerr background.
Such a calculation will probably be a very significant step towards an understanding of the range of validity 
of the Cosmic Censorship Conjecture.

If, however, we assume that $\delta b_{\rm ph}$ decreases in the almost-extremal limit under the effect of the gravitational self-force,
a change of order $\delta b_{\rm ph}={\cal O}(\epsilon)$ may also be enough to prevent the ingoing fluxes from
falling into the horizon and create a naked singularity.
Suppose in fact that $\delta b_{\rm ph}$ were negative enough to prevent ultrarelativistic particles 
with impact parameter $b$ in the JS range ($b=2+\delta\epsilon$ with $2 \sqrt{2}<\delta<4$, \textit{cf.}~Sec.~\ref{orbits}) 
from falling into the horizon. These
particles will then describe a large number of quasi-circular orbits near the periastron before
being scattered. During these quasi-circular orbits, they will emit gravitational
fluxes  $E_{\rm rad}$ and $L_{\rm rad}$ into the horizon, with $L_{\rm rad}/E_{\rm rad}=1/\Omega_{\rm peri}=b$,
and these fluxes tend to spin the black hole up, possibly above the Kerr limit [\textit{cf.}~discussion around Eq.~\eqref{intermediate_spin}].
However, it is well-known that in the eikonal limit (\textit{i.e.,}~in the small-wavelength limit, 
corresponding to large $m$'s in the decomposition of Secs.~\ref{numerics} 
and~\ref{extrapolation}) gravitational waves behave like massless particles (``gravitons''), their propagation in a Kerr spacetime being 
regulated by a radial effective potential that is the same as the radial effective potential regulating the 
motion of photons [\textit{cf.}~Eq.~\eqref{rdot}].\footnote{The eikonal approximation is a 
very good one for the gravitational waves emitted by the JS orbits because,
as we stressed in Sec.~\ref{extrapolation}, the fluxes in the large-$m$ multipole moments turn out to be very important for these orbits.}
We can therefore think of the ingoing fluxes as being made of  wavepackets with impact parameter $b_{\rm g}$ ranging from $0$ to $+\infty$,
but with a distribution centered (roughly) on the impact parameter $b$ of the particle producing them (because the ratio of the total
ingoing fluxes must be $L_{\rm rad}/E_{\rm rad}=b$). Clearly, the only wavepackets capable of overspinning the black hole are
those with $b_{\rm g}$ in the JS range $b_{\rm g}=2+\delta\epsilon$ with $2 \sqrt{2}<\delta<4$. However, if the effective potential 
for ultrarelativistic orbits (and therefore for gravitons) changes under the effect of the self-force in such a way
as to prevent the JS orbits from being captured [\textit{i.e.}~if $\delta b_{\rm ph}={\cal O}(\epsilon)$ is sufficiently negative],
it will also scatter the  gravitational wavepackets with $b_{\rm g}$ in the JS range.

We stress that Jacobson and Sotiriou proposed also a different scenario in which naked singularities might form. In particular,
they consider a \textit{spinning} particle having energy $E$ and spin $S$, with the energy satisfying the limits \eqref{limitsE}
and the spin satisfying the limits  \eqref{limitsJ} (where we identify $L$ with the particle's spin $S$). They also assume that the particle
has spin parallel to the black-hole spin and that it is dropped into the black-hole horizon along the common direction of their spins. The final spin
$a_f^{\rm JS}$ of the resulting
black hole will be larger than 1, because the lower bound in Eq.~\eqref{limitsJ} is achieved precisely by imposing $a_f^{\rm JS}>1$.
Also, the condition for a spinning particle to fall into a Kerr black hole is $E>\Omega_H S$~\cite{Jacobson:2009kt}, 
where $\Omega_H\approx 1/2-\epsilon$ is the horizon frequency, and the upper bound of Eq.~\eqref{limitsJ} can 
indeed be written in that form, \textit{i.e.}
the upper bound of Eq.~\eqref{limitsJ} ensures that the spinning particle actually falls into the black-hole horizon.

From the discussion above, however, it is clear that the conservative self-force will modify the metric and therefore the
horizon frequency by terms of order $E\sim \epsilon$. In particular, if the horizon frequency 
\textit{increases} under the effect of the  conservative self-force (just like the ISCO frequency increases in a Schwarzschild spacetime~\cite{barack_sago}) 
and becomes $\tilde{\Omega}_H=\Omega_H+\kappa \epsilon
\approx 1/2-(1-\kappa)\epsilon$ (with $\kappa>0$ being a coefficient),  Eq.~\eqref{limitsJ} would become
\beq
2 \epsilon^2 + 2 E + E^2<S<  [2 + 4 (1-\kappa) \epsilon] E\,.\label{limitsJ_modified}
\eeq
Imposing that the upper bound in this equation be larger than the lower bound [\textit{i.e.,} imposing 
that the interval described by Eq.~\eqref{limitsJ_modified} is not empty], we immediately obtain that if $\kappa>1-1/\sqrt{2}\approx0.293$
there are \textit{no} orbits that both fall into the black hole and create a naked singularity. In other words, the conservative
self-force may be enough to prevent the JS from forming naked singularities even in the case of particles with spin. 

Finally, proceeding exactly in the same
way as we just did for spinning particles, it possible to show that the self-force may also be enough to prevent naked singularities from forming
in the case of non-spinning particles on \textit{bound} JS orbits. As we stressed in Sec.~\ref{orbits}, these orbits start 
from very close to the black-hole horizon and do not orbit the light ring multiple times, so the analysis of the previous sections does not apply to them.
However, because their energy and angular momentum must still satisfy Eqs.~\eqref{limitsJ} and \eqref{limitsE}, 
the horizon frequency will change
by ${\cal O}(\epsilon)$ due to the conservative self-force. In particular, if the horizon's frequency increases, then the upper bound in  Eq.~\eqref{limitsJ}
will be lowered and the allowed angular momentum interval might possibly shrink to nothing. 

\section{Conclusions}\label{concl}
As discussed in the introduction, there is strong circumstantial evidence for the stability of Kerr black holes, which leads us to believe 
that rumors of their death may have again been greatly exaggerated. Nevertheless, the particular mechanism proposed by Jacobson and Sotiriou~\cite{Jacobson:2009kt}
is exciting enough to deserve careful thought. The mechanism can be at play in astrophysical settings and the
understanding of why it fails (or not)
will certainly shed light on highly dynamical processes close to extremal Kerr black holes. We have shown 
that radiation by point particles close to the last stable circular geodesic is suppressed as the black hole 
approaches extremality, providing further support to earlier results by 
Chrzanowski and Misner \cite{Chrzanowski:1976jy,Chrzanowski:1974nr}. 
Indeed, we have shown that if one modifies Jacobson and Sotiriou's analysis by replacing the test particle with a ring of particles, 
the gravitational radiation is suppressed by interference effects.
It therefore seems that the conservative self force  
might be the main effect preventing violations of the Cosmic Censorship Conjecture.
While we have provided arguments in favor of this picture, a rigorous proof is still unavailable,
and the role of the conservative self-force in the Cosmic Censorship Conjecture remains
 an outstanding open issue.

\vspace{0.2cm}
\begin{acknowledgments}
  This work was supported by the {\it DyBHo--256667} ERC Starting
  Grant and by FCT - Portugal through projects PTDC/FIS/098025/2008,
  PTDC/FIS/098032/2008 CTE-AST/098034/2008 and CERN/FP/116341/2010.
  G. K. acknowledges research support from NSF grants PHY-0902026 and PHY-1016906.
E. B. acknowledges support from NSF grant PHY-0903631.
 Most of the numerical simulations needed for this work were performed on Georgia Tech's 
  Keeneland supercomputer under project number UT-NTNL0036.  
\end{acknowledgments}


\end{document}